\documentclass[pdftex,twocolumn,epjc3]{svjour3}          

\RequirePackage[T1]{fontenc}

\smartqed  

\RequirePackage{graphicx}
\RequirePackage{mathptmx}      
\RequirePackage{flushend}
\RequirePackage[numbers,sort&compress]{natbib}
\RequirePackage[colorlinks,citecolor=blue,urlcolor=blue,linkcolor=blue]{hyperref}

\journalname{Eur. Phys. J. C}

\begin{document}

\title{Search for Weakly Interacting Sub-eV Particles with the OSQAR Laser-based Experiment: Results and Perspectives
}


\author{P. Pugnat \thanksref{e1,lncmi1,lncmi2}
\and R.Ballou \thanksref{neel1,neel2}
\and M. Schott \thanksref{univmainz}
\and T. Husek \thanksref{charlesuniv}
\and M. Sulc \thanksref{univliberec}
\and G. Deferne \thanksref{cern}
\and L. Duvillaret \thanksref{imep}
\and M. Finger Jr. \thanksref{charlesuniv}
\and M. Finger \thanksref{charlesuniv}
\and L. Flekova \thanksref{czech}
\and J. Hosek \thanksref{czech}
\and V. Jary \thanksref{czech}
\and R. Jost \thanksref{liphy1,liphy2}
\and M. Kral \thanksref{czech}
\and S. Kunc \thanksref{univliberec}
\and K. Macuchova \thanksref{czech}
\and K. A. Meissner \thanksref{itppol}
\and J. Morville \thanksref{lasim1,lasim2}
\and D. Romanini \thanksref{liphy1,liphy2}
\and A. Siemko \thanksref{cern}
\and M. Slunecka \thanksref{charlesuniv}
\and G. Vitrant \thanksref{imep}
\and J. Zicha \thanksref{czech}
}

\thankstext{e1}{e-mail: pierre.pugnat@lncmi.cnrs.fr}

\institute{CNRS, LNCMI, F-38042 Grenoble, France\label{lncmi1}
\and Univ. Grenoble Alpes, LNCMI, F-38042 Grenoble, France\label{lncmi2}
\and CNRS, Institut N\'eel, F-38042 Grenoble, France\label{neel1}
\and Univ. Grenoble Alpes, Institut N\'eel, F-38042 Grenoble, France\label{neel2}
\and University of Mainz, Institute of Physics, 55128 Mainz, Germany\label{univmainz}
\and Charles University, Faculty of Mathematics and Physics, Prague, Czech Republic\label{charlesuniv}
\and Technical University of Liberec, 46117 Liberec, Czech Republic\label{univliberec}
\and CERN, CH-1211 Geneva-23, Switzerland\label{cern}
\and Grenoble INP - Minatec \& CNRS, IMEP-LAHC, F-38016 Grenoble, France\label{imep}
\and Czech Technical University, Prague, Czech Republic\label{czech}
\and Univ. Grenoble Alpes, LIPhy, F-38000 Grenoble, France\label{liphy1}
\and CNRS, LIPhy, F-38000 Grenoble, France\label{liphy2}
\and University of Warsaw, Institute of Theoretical Physics, 00-681 Warsaw, Poland\label{itppol}
\and Univ. Claude Bernard Lyon-1, Institut Lumi\`ere Mati\`ere, F-69622 Villeurbanne, France\label{lasim1}
\and CNRS, Institut Lumi\`ere Mati\`ere, F-69622 Villeurbanne, France\label{lasim2}
}

\date{Received: date / Accepted: date}

\maketitle

\begin{abstract}

Recent theoretical and experimental studies highlight the possibility of new fundamental particle physics beyond the Standard Model that can be probed by sub-eV energy experiments. The OSQAR photon regeneration experiment looks for \textquotedblleft~Light Shining through a Wall~\textquotedblright~(LSW) from the quantum oscillation of optical photons into \textquotedblleft~Weakly Interacting Sub-eV Particles~\textquotedblright~(WISPs), like axion or axion-like particles~(ALPs), in a 9~T transverse magnetic field over the unprecedented length of $2 \times 14.3$~m. No excess of events has been detected over the background. The di-photon couplings of possible new light scalar and pseudo-scalar particles can be constrained in the massless limit to be less than $8.0\times10^{-8}$ ~GeV$^{-1}$. These results are very close to the most stringent laboratory constraints obtained for the coupling of ALPs to two photons. Plans for further improving the sensitivity of the OSQAR experiment are presented. 

\end{abstract}

\section{Introduction}
\label{intro}

Particle and astroparticle physics beyond the Standard Model (SM) is not restricted to the high energy frontier. Many extensions of the SM, in particular those based on supergravity or superstrings, predict not only massive particles like the Weakly Interacting Massive Particles (WIMPs), but also Weakly Interacting Sub-eV Particles (WISPs). 
WIMPs can be searched for at TeV colliders such as the Large Hadron Collider (LHC) at CERN. 
WISPs in contrast seem most likely to be detected in low energy experiments based on lasers, microwave cavities, strong electromagnetic fields or torsion balances \cite{JaRi2010,AWGr2011}. As the first and paradigmatic example of WISPs, there is the axion \cite{Weinberg1978,Wilczek1978}. Axion remains one of the most plausible solutions to the strong-CP problem \cite{Peccei1977}. It also constitutes a fundamental output of the string theory in which a number of axions or axion like particles (ALPs) are predicted without any contrivance \cite{Svrek2006,Cicoli2012}. The interest for axion search extends beyond particle physics, since such a hypothetical light spin-zero particle is considered as a serious non-supersymmetric dark-matter candidate \cite{Bradlay2003,Arias2012a,Arias2012b}. Axions might also explain a number of strange astrophysical phenomena, such as the universe transparency to very high energy photons (> 100 GeV) \cite{Meyer2013}, the anomalous white dwarf cooling \cite{Melendez2012a,Melendez2012b} or the recently discovered gamma ray excesses in galaxy clusters \cite{Cicoli2014}.

The OSQAR experiment is at the forefront of this emerging low energy frontier of particle/astroparticle physics. It combines the simultaneous use of high magnetic fields with laser beams in two distinct experiments. In the first one, it looks for the photon regeneration effect as a Light Shining through the Wall (LSW), whereas in the second one, it aims for a first measurement of ultra-fine vacuum magnetic birefringence, either associated to virtual WISPs or merely the one predicted in Quantum Electro-Dynamics, and vacuum magnetic dichroism, associated to real WISPs \cite{Pugnat2010,Sulc2013}. This article focuses on the OSQAR LSW experiment and is in the same line as the pioneering work in the early $1990$ that excluded ALPs with a di-photon coupling constant $g_{A\gamma\gamma} > 7\times10^{-7}$~GeV$^{-1}$ for an ALP mass below 1~meV \cite{Cameron1993}. A strong revival of interest in this type of experiment has been triggered by the announcement in $2006$ of possible positive ALP signal \cite{Zavattini2006}, but all recent LSW worldwide experiments have excluded such a possibility \cite{Robilliard2007,Chou2008,Afanasev2008,Zavattini2008,Pugnat2008,Ehret2010} and the present most stringent limit, obtained by the ALPS collaboration, excludes ALPs with $g_{A\gamma\gamma}  > 6.5 \times10^{-8}$~GeV$^{-1}$ in the massless limit  \cite{Ehret2010}. We report below on the last OSQAR results, approaching this exclusion limit for ALP search.

\section{The OSQAR Photon Regeneration Experiment}
\label{OSQARPRExp}

\begin{figure}
\includegraphics[keepaspectratio=true, width=8.5cm]{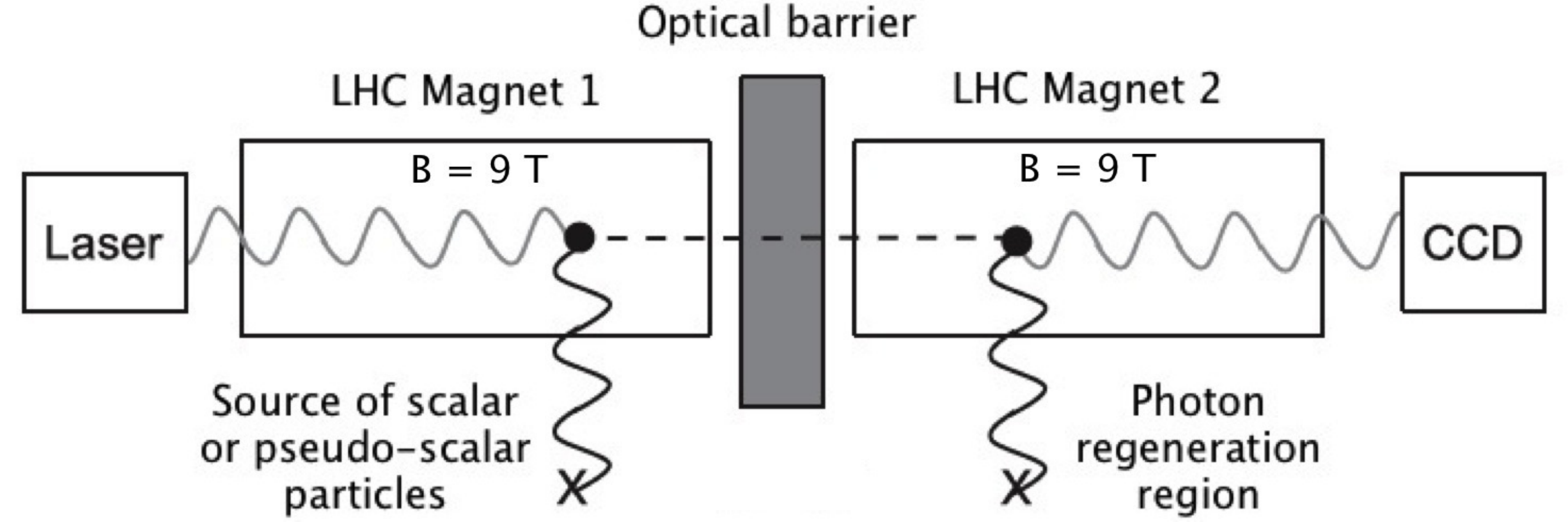}
\caption{Schematic view of the OSQAR photon regeneration experiment using two spare dipole magnets of the Large Hadron Collider (LHC), each providing a transverse magnetic field of $9~T$ over $14.3~m$.}
\label{fig:osqarexp}
\end{figure}

LSW experiment is the simplest and most unambiguous laboratory experiment to look for WISPs. It combines photon-to-WISP and WISP-to-photon double quantum oscillation effect, possibly induced  in a transverse magnetic field, with the weakness of the coupling of WISPs to matter \cite{Sikivie1983,vanKibber1987,Arias2010}. When a linearly polarized laser light beam propagating in a transverse magnetic field for instance is sent through an optical barrier only the photons converted to WISPs due to the mixing effect will not be absorbed. These WISPs will propagate freely through the barrier owing to the weak coupling to matter before being reconverted in photons of same energy that can be easily detected and identified (Figure \ref{fig:osqarexp}).

In the case where the WISP is an ALP, the photon-to-ALP $(\gamma \rightarrow A)$ conversion probability, as well as the ALP-to-photon $(A \rightarrow \gamma)$ one, in vacuum over a length L permeated by a transverse magnetic field B are given by \cite{Sikivie1983,vanKibber1987,Arias2010}:
\begin{equation}
\label{equ:photonWISPpc}
P_{\gamma \leftrightarrow A} = \frac{1}{4} (g_{A\gamma\gamma}  BL)^2 \left( \frac{2}{qL} \sin\frac{qL}{2} \right)^2
\end{equation}
using Heaviside-Lorentz units ($\hbar=c=1$). $g_{A\gamma\gamma}$ is the ALP di-photon coupling constant and $q = \|k_\gamma - k_A\|$ the momentum transfer, with $k_ A = (\omega^2-m_A^2)^{1/2}$ and $k_\gamma=\omega$ where $\omega$ is the energy for photons and ALPs and $m_A$ is the ALPs mass. The form factor of the conversion probability is at maximum for qL = 0, which corresponds to the limit $m_A \ll \omega$ otherwise incoherent effects of the $\gamma \leftrightarrow A$ oscillation reduce the conversion probability. With pseudo-scalar ALPs, which includes the axion, (resp. with scalar ALPs) the probability is maximum when the linear polarization of light is parallel (resp. perpendicular) to the magnetic field and vanishes when this polarization is rotated in the perpendicular direction. The WISPs other than the ALPs, with conversion probability independent of the magnetic field or depending differently from it, were ignored. Owing to the weak initial photon flux our data were not competing with those reported in the literature \cite{Ehret2010}.

The photon flux after the optical barrier (Figure \ref{fig:osqarexp}) is given by:
\begin{equation}
\label{equ:photonflux}
\frac{dN_\gamma}{dt} = \frac{P}{\omega} \eta~P_{\gamma\leftrightarrow A}^2
\end{equation}
where $P$ is the optical power, $\omega$ the photon energy and $\eta$ the efficiency of the detector. Equation \ref{equ:photonflux} shows that the regenerated photon flux scales as $ (BL)^4$, highlighting the interest of using high field magnets over the longest optical path length. With this respect, the LHC dipole magnets cooled down to 1.9~K with superfluid He and each of them being able to produce a transverse magnetic field of 9~T over 14.3~m constitute nowadays the state of the art.

The present experimental setup of the OSQAR photon regeneration experiment is using two LHC dipole magnets separated by an optical barrier as shown in Figure \ref{fig:osqarexp}. Each dipole magnet is connected to a cryogenics feedbox. The total length of the experiment is about $53~m$. Magnet apertures is pumped typically down to $10^{-6} - 10^{-7}$~mbar with two turbomolecular pumping groups. An ionized Ar$^+$ laser has been used to deliver in multi-line mode $3.31 \pm 0.03$~W of optical power in average with approximately $2/3$ of the optical power at 514~nm ~(2.41~eV) and $1/3$ at 488~nm~(2.54~eV). Its beam has a well defined linear polarization parallel to the magnetic field optimized for the search of new pseudoscalar/axion particles. To look for scalar particles, a $\lambda / 2$ wave plate with antireflective coating layers is oriented at $45^\circ$ and inserted at the laser exit to align the polarization perpendicularly to the magnetic field. The laser beam divergence has been reduced with a specially developed beam expander telescope. By combining converging and diverging antireflective coated lenses, the optical Gaussian beam with a waist equal to 0.708~mm at the level of the output coupler, was shaped to obtain a spot size not exceeding 6~mm of diameter at 53~m of distance. To minimize spherical aberrations, lenses were mounted with planar surfaces face-to-face and thermal effects on lenses were minimized by developing proper supports. For photon counting, a liquid nitrogen (LN$_2$) cooled CCD detector from Princeton Instrument (model LN/CCD-1024E/1) has been used. The CCD chip of $26.6\times 6.7$~mm$^2$ is composed by a 2D array of $1024 \times 256$ square pixels of $26 ~\mu$m size. The quantum efficiency of the detector is equal to 30\% (including the gain factor of the CCD) for the Ar$^+$ laser wavelengths. Dark current and readout noise at 20~kHz are typically equal to 0.5~e$^{-}$/pixel/hour and 3.4~e$^{-}$rms respectively. An optical lens with a focal length of 100~mm has been installed just in front of the detector to focus the laser beam on the CCD chip.

\section{Preparatory Phase, Experimental Runs and Data Taking}

The two LHC dipole magnets used for OSQAR have been installed on their horizontal cryogenic bench and precisely aligned with a Laser Tracker LTD 500 from Leica before being thoroughly tested at 1.9~K and used in routine operation to provide the 9~T magnetic field over $2 \times 14.3$~m. In particular, the field strength and field errors of both LHC dipole magnets used have been precisely characterized. The long term stability of the alignment of the laser beam traveling through the aperture of both LHC dipole magnets has been also carefully checked and controlled periodically with the LN$_2$ cooled CCD.

\begin{figure}
\includegraphics[keepaspectratio=true, width=8cm]{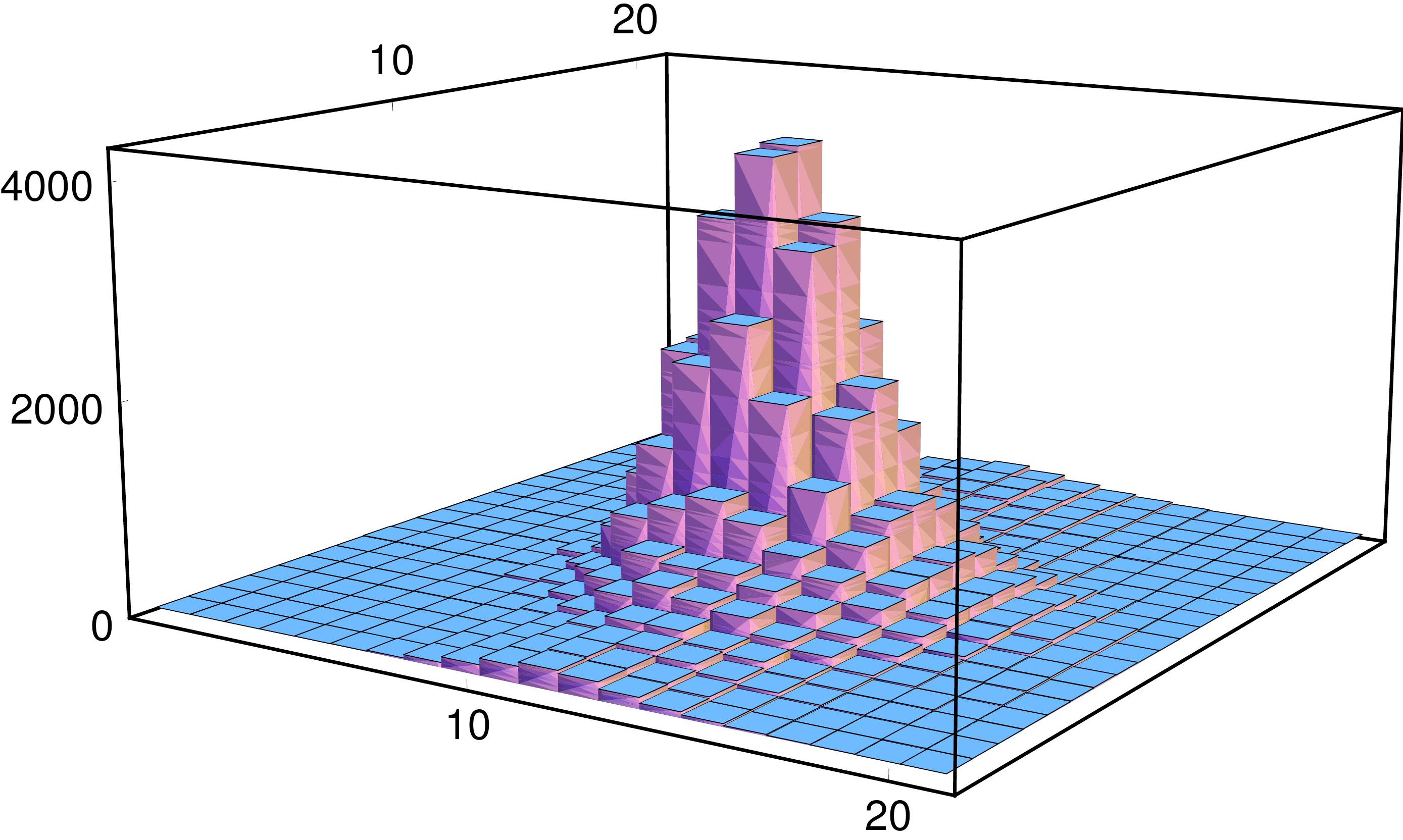}
\caption{Focused and attenuated optical beam signal recorded during the pointing of the laser beam on the CCD. Each square represents a super-pixel consisting of 4 CCD pixels ($2\times 2$ binned). The height corresponds to the photon count.}
\label{fig:osqarInBeamSpot}
\end{figure}

A typical experimental run starts and ends with laser beam alignments using absorptive filters to reduce the optical intensity below the saturation level of the CCD. An example of raw signals recorded with the CCD in 2D mode is given in Figure \ref{fig:osqarInBeamSpot}. The photons regenerated from ALPs can be efficiently and unambiguously identified from the shape and the precise localization of the known expected signal. The CCD pixels were $2 \times 2$-binned at the read out step, {\it i.e.} the charge from four neighbor pixels (2 along each direction of the array) are combined into the charge of a single super-pixel. The CCD spectra then is composed of a 2D array of $512 \times 128$ counted entries. The central peak corresponding to the focused and attenuated laser beam has a much broader width compared to parasitic signals such as those coming from cosmic rays. Table~\ref{exppar} gives a summary of the duration of data taking as a function of various configurations of the experiment including the effective optical power,  {\it i.e.} after subtracting all losses due to parasitic reflexions of light.

\begin{table}[h]
\centering
\caption{Summary of the characteristics of the experimental runs based on $900~s$ exposure time frames.}
\label{exppar}
\begin{tabular*}{\columnwidth}{@{\extracolsep{\fill}}cccc@{}}
\hline
Polarization & Magnetic  & Laser  & Acquisition \\  Angle & Field & Power & Total Time\\ \hline $0^\circ$ (Pseudoscalar) & 9 T & 1.62 W & 21.75 hours    \\ \hline $90^\circ$ (Scalar) & 9 T & 1.64 W & 23.75 hours \\ \hline  (Background) & 9 T & 0 W & 4 hours \\\hline
\end{tabular*}
\end{table}

\section{Data Analysis and New Exclusion Limits}

Cosmic rays contamination were corrected by filtering the data against spatially sharp signals, by comparing counts between neighboring super-pixels. This correction has been validated on the laser beam profile data to ensure that it does not impact some broader signal distributions compatible with a regenerated photon beam. As a matter of fact, in order to test the correction on a spectrum which would be compared with a spectrum that might show a signal, the laser beam data was rescaled to have maximum pixel intensity of the order of ten to twenty times the standard deviation of uncontaminated background noise then was added to a background spectrum acquired by switching off the laser. This cosmic cleaning procedure was applied systematically to each 2D recorded spectrum of 900~s exposure time. 

We observed a constant count-rate of 0.0762(5)~s$^{-1}$ in each super-pixel due to the used electronic readout in runs with and without the laser beam. This constant count-rate and the associated uncertainty are computed as the average and standard deviation of the counts over the super-pixels (i.e. the $2 \times 2$-binned pixels), after correction of the cosmic rays contamination. It displayed excellent stability in time, fluctuating by an amount less than 0.0001~s$^{-1}$ between consecutive recorded spectra when the computed uncertainty for each recorded spectrum is around 0.0005~s$^{-1}$. Even when it slightly evolved in time, for instance after cryogenic refilling, the change for consecutive recorded spectra was always clearly smaller than the mean uncertainty. We are legitimated to ignore these weak drifts by subtracting from each recorded spectrum the average count computed after correction of the cosmic rays contamination, before summation to produce a final spectrum of accumulated counts. This then consists of negative as well as positive super-pixel counts with zero average. It is important to emphasize that only the width of these count rate fluctuations limit the sensibility of the experiment and not the count rate itself. Alternatively, the subtraction of the background acquired by switching off the laser, after correction of the cosmic rays contamination, to a spectrum for an ALP search, that immediately followed or preceded, gives no significant difference. 

We furthermore observed and carefully confirmed that systematically the background was spatially flat over a wide surface of the CCD. We used this property to collect and treat the data in terms of surface distribution of counts, with a region of expected signal surrounded by background regions of equal size, which thus are simultaneously measured. This avoids having to perform a background count with the laser switched off after each recording of a spectrum for an ALP search, which {\it \`a priori} would have been necessary in order to care about possible long time drifts but would have been uselessly time consuming. We of course checked before and after each run that the position of the expected signal on the CCD was spatially stable. Although not systematically, we nevertheless periodically collected background data of 900~s exposure time with the laser beam switched off to ensure that the CCD was running properly, with counts on super-pixels showing similar statistics up to cosmic rays contamination, and to preserve the possibility to subtract these background spectra to the spectra for the ALP search.

\begin{figure}
\includegraphics[keepaspectratio=true, width=8.5cm]{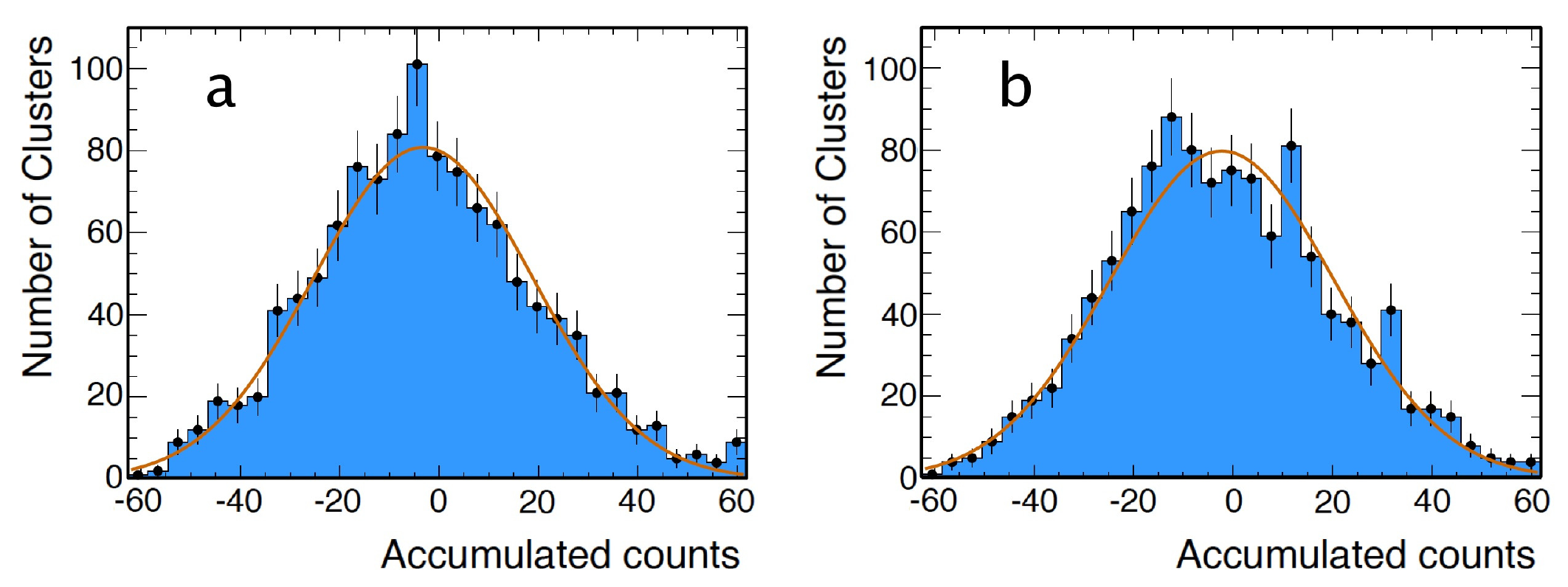}
\caption{Histograms of the accumulated counts from clusters of $4 \times 5$ super-pixels of the CCD for the search of a) pseudoscalar particles, with total time integration of 21.75 hours, and b) scalar particles, with total time integration of 23.75 hours.}
\label{fig:osqarhistogram}
\end{figure}

\begin{figure}
\includegraphics[keepaspectratio=true, width=8.5cm]{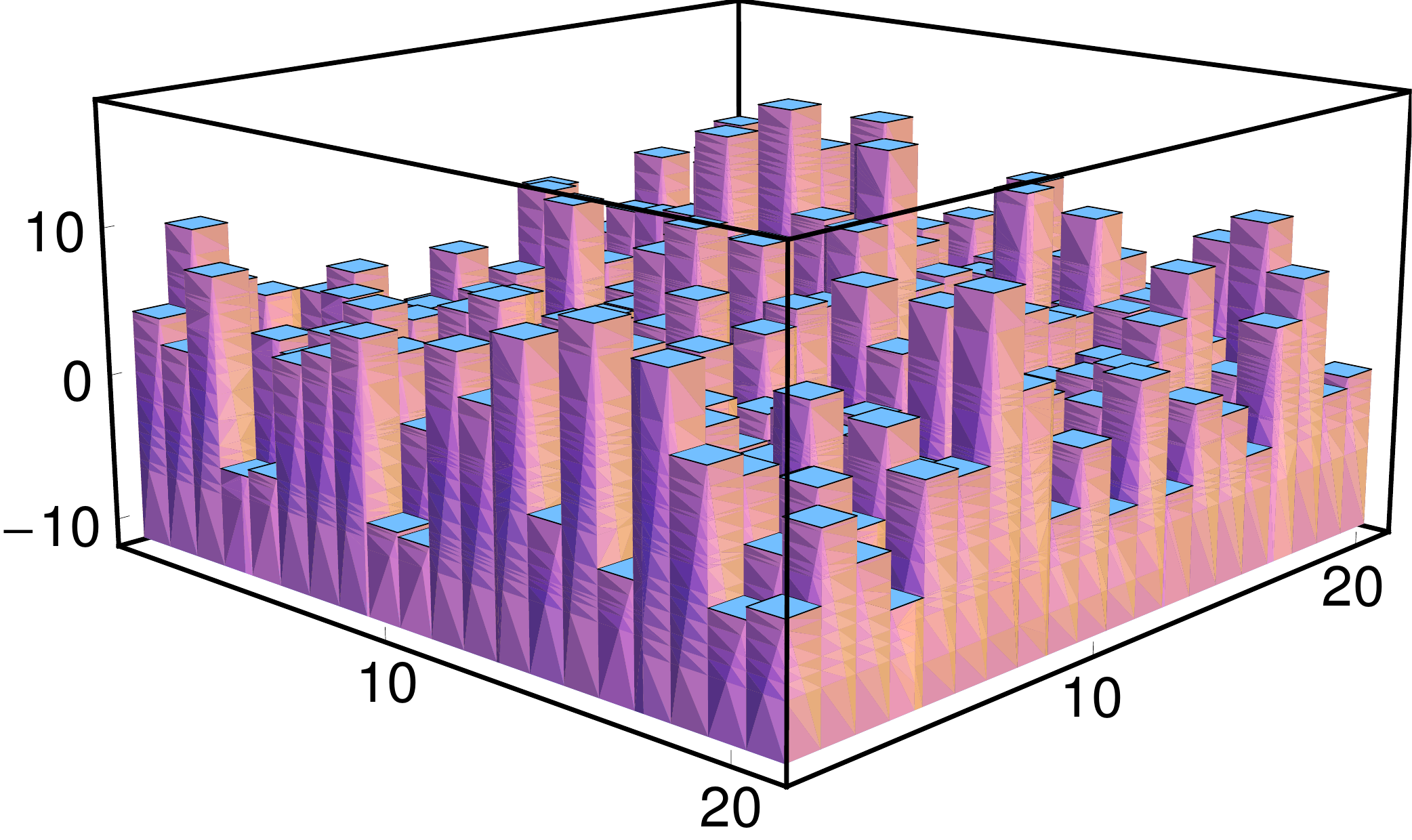}
\caption{Simulated signal as it would have been observed for a WISP-diphoton coupling of $12 \times 10^{-8}$~GeV$^{-1}$, which correspond to a signal that is five times larger than the width of the background fluctuations.}
\label{fig:osqarvisualexpected}
\end{figure}

After correction of the cosmic rays contamination, checking of the background (drift and flatness) and subtraction of the constant background, the spectra corresponding to the search of either pseudoscalar  or scalar particles have been added separately, leading to two final spectra of accumulated counts. As a last step, each final data spectrum has been clustered to optimize the experimental sensitivity with respect to the detection of the photon $\leftrightarrow$ ALP conversions. On increasing the cluster size this sensitivity is decreased by the increase of the fluctuations of the cluster counts, but is increased by the proportion of optical power integrated on the cluster of expected signal. The optimal size of the clusters corresponds to a rectangle of $4 \times 5$ super-pixels and contains $30.3~\%$ of the signal in the pointing zone of the laser beam. One then isolates 709 contiguous clusters of $4 \times 5$ super-pixels, namely a central cluster at which the signal should be looked at and 708 clusters available for statistical analysis of the background noise. The integrated count of each cluster is defined by the sum of the accumulated counts of the corresponding $4 \times 5 = 20$ super-pixels. We show in Figure \ref{fig:osqarhistogram} the histograms of this integrated signal over all the clusters. The distributions can be accurately described by a gaussian function the width of which determines the sensitivity of the experiment. The best fitted parameters are listed in Table~\ref{expres}. Note that the observed Gaussian HWHM is close to the expectation value, in the range of $20-21$ accumulated counts, derived from the CCD characteristics given in Section \ref{OSQARPRExp} taking into account the CCD area covered by the 709 clusters selected for the data analysis. Within this picture we are able to claim that a signal is detected at a 95\% confidence level if an excess of counts larger than two times the standard deviation of the gaussian distribution is recorded in the cluster at which the laser beam pointed. A simulated signal as it would have been observed for an excess of counts five times larger is displayed in Figure \ref{fig:osqarvisualexpected}. It was obtained by adding to the spectrum of accumulated counts the appropriately rescaled signal of the beam displayed in Figure \ref{fig:osqarInBeamSpot}. It visually disappears in the noise at the limit of the width of the fluctuations of the cluster counts.

\begin{table}
\centering
\caption{Parameters of Gaussian fit  of histograms (Figure \ref{fig:osqarhistogram}) and deduced sensitivity (HWHM stands for Half Width at Half Maximum). The goodness of fits, $\chi^2$ /  (number of degrees of freedom), is 1.3 for both pseudoscalar and scalar particle searches.}
\label{expres}
\begin{tabular*}{\columnwidth}{@{\extracolsep{\fill}}cccc@{}}
\hline
\multicolumn{1}{@{}l}{} &  Gaussian  &  Gaussian &  Sensitivity \\  & Center & HWHM & (photons / s) \\\hline Pseudoscalar & $-0.96 \pm 0.9$ & $21.1\pm 0.9$ &  $1.76 \times 10^{-3}$  \\\hline Scalar  & $-1.56 \pm 1.7$ & $22.7\pm 1.8$ & $1.71 \times 10^{-3}$  \\\hline
\end{tabular*}
\end{table}

\begin{figure}
\includegraphics[keepaspectratio=true, width=8.5cm]{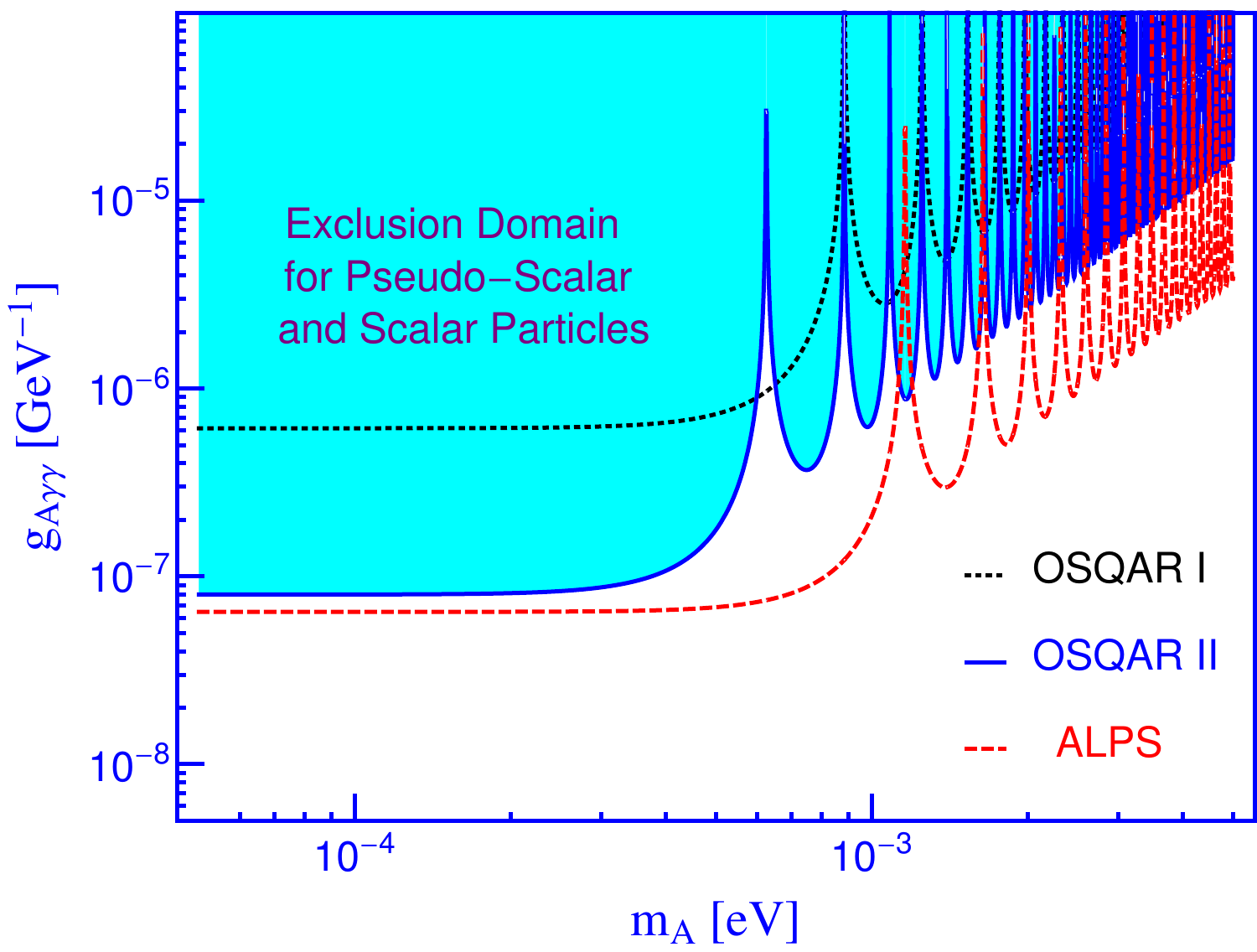}
\caption{Exclusion limits from the pseudo-scalar and scalar particle searches for the last two runs of the OSQAR experiment. The last result  in vacuum of the ALPS collaboration is also shown in dashed red.}
\label{fig:osqarexclusionlimits} 
\end{figure}

No excess of accumulated counts can be detected for the cluster located at the pointing zone of the laser beam  for the search of either pseudoscalar or scalar particles. Conservatively the $95 \%$ confidence interval for pseudoscalar particle can be set from Gaussian fitting parameters of Table~\ref{expres} to 41 cluster counts. The predicted CCD quantum efficiency of 30\% corresponds to 137 cluster counts and gives a limit to the sensitivity of 137 photons/21.75 hours = 6.32 photons/hour or $1.76 \times 10^{-3}$~photons/s. A very close sensitivity is obtained for scalar particle search and both these results are reported in Table~\ref{expres}. 
Inserting this photon flux into Equation \ref{equ:photonflux}, the exclusion limits for ALPs parameters can be determined and are compared in Figure \ref{fig:osqarexclusionlimits} to the last results obtained in vacuum by the ALPS collaboration \cite{Ehret2010}. 

The di-photon coupling $g_{A\gamma\gamma}$ is less constrained when the mass $m_A$ is larger than $5 \times 10^{-3}$~eV. It is even unconstrained for specific values of the mass $m_A$. This emanates from the loss of coherence of the $\gamma\leftrightarrow A$ oscillations and occurs when the product qL of the momentum transfer q and spatial length L permeated by the magnetic field is an integer multiple of 2$\pi$: $qL = 2 u \pi$ where $u$ is a strictly positive integer (cf. equation \ref{equ:photonWISPpc}). The condition is met for the mass $m_A^\ast (u) = \omega \sqrt{1-(1-2u\pi/\omega L)^2} \approx \sqrt{4u\pi\omega/L}$. We notice in Figure \ref{fig:osqarexclusionlimits} that $m_A^\ast (u)$ for the OSQAR I experiment is equal to $m_A^\ast (2u)$ for the OSQAR II experiment. This merely accounts for the use of one dipole in OSQAR I and two dipoles in OSQAR II, so that the spatial length L permeated by the magnetic field was doubled in this second experiment. The ALPS experiment used a HERA dipole providing a magnetic field of 5~T in a length of 8.8~m and photons of a different energy $\omega$ \cite{Ehret2010}. These gaps in the exclusion limits can be filled by introducing a gas in the medium permeated by the magnetic field to shift the photon momentum $k_\gamma$ and thus the momentum transfer q associated with $\gamma\leftrightarrow A$ oscillations \cite{Pugnat2008,Ehret2010}.

The values of coupling constants of possible new light pseudo-scalar and scalar particles that can couple to two photons is constrained in the massless limit to be less than $8.0 \times 10^{-8}$~GeV$^{-1}$. This result constitutes a gain by one order of magnitude in sensitivity with respect to our previous investigation \cite{Pugnat2008} and the first precise confirmation of the present most stringent exclusion limit for ALPs obtained from a purely laboratory experiment \cite{Ehret2010}. The accuracy of our results might have been improved by increasing the number of recorded spectra, but our laser source unfortunately broke down. It is emphasized besides, that the di-photon coupling constant g$_{A\gamma\gamma}$ scales only with minus one-eighth power of the exposure time $t$ (g$_{A\gamma\gamma} \sim  t^{-1/8}$) and improving the results by a factor of about 1.25, to exceed the exclusion limit in the massless limit of the ALPs experiment, would have required to record 10 times more spectra.

\section{Anticipated improvements and awaited results}

The next steps of the OSQAR LSW experiment will be based on three main improvements. First, the axion source will be amplified by upgrading the optical power. Second, the photo-regeneration will be resonantly enhanced by implementing the setup with two optical resonators placed before and after the optical barrier and phase-locked \cite{Hoogeveen1991,Sikivie2007}. Third, the possibility of further increasing the number of LHC dipoles is currently studied to augment the $\gamma \leftrightarrow A$ conversion probabilities through the field integral BL, starting from a configuration using $2+2$ dipoles for the coming years.

\begin{figure}
\includegraphics[keepaspectratio=true, width=8.5cm]{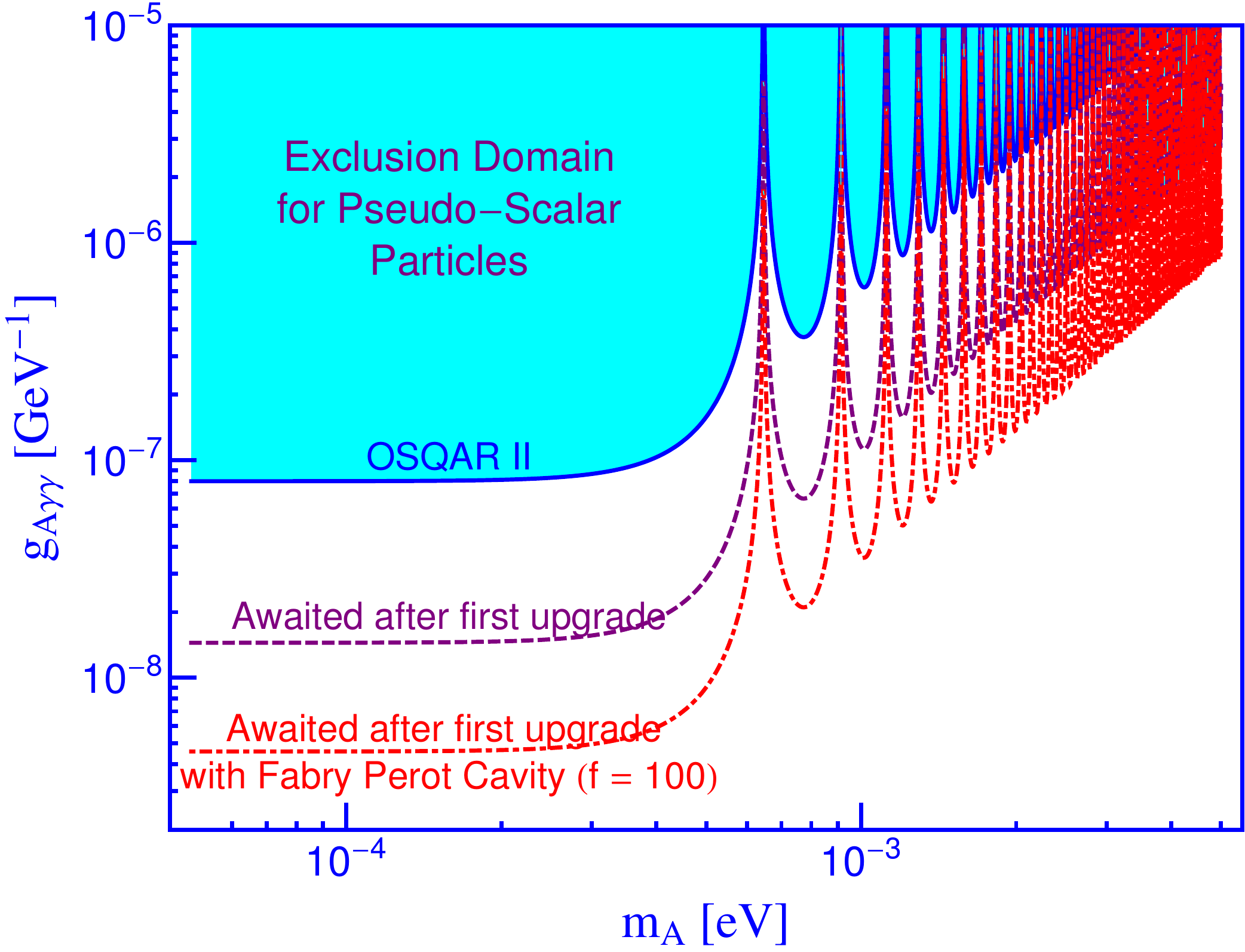}
\caption{Exclusion limits awaited from the upgrade of the optical power and photon detection and additional improvement awaited from insertion of the first magnet in a Fabry-Perrot cavity of finesse f = 100.}
\label{fig:osqarawaited}
\end{figure}

The optical power will be upgraded with a more powerful laser source of 15~W and by adding a Fabry-Perot cavity, as what was carried out before by the ALPS experiment \cite{Ehret2010}. We have in fact recently managed to phase-lock to a laser source a cavity of length l = 20~m and finesse f = 100, that can be inserted in the bore of a first dipole magnet. The detection system also will be optimized thanks to a new CCD with improved properties, in particular a  quantum efficiency of 95\%. The laser beam finally will be focussed over a smaller number of pixels which will be binned at the read out step to form a single super-pixel. It was indeed observed in the present experiment that the laser beam pointing was  stable, without shift beyond a super-pixel over the recording time of a spectrum. It also turns out that the expected signal thus focussed can be distinguished still easily from possible cosmic rays contamination, since the probability that this spoils the super-pixel at which the signal is awaited is extremely weak and in the scarce case where this might occur an unphysically large signal will be observed. We show in figure \ref{fig:osqarawaited} the exclusion limits awaited from these anticipated upgrade of optical power and optimization of detection. 

\begin{figure}
\includegraphics[keepaspectratio=true, width=8.5cm]{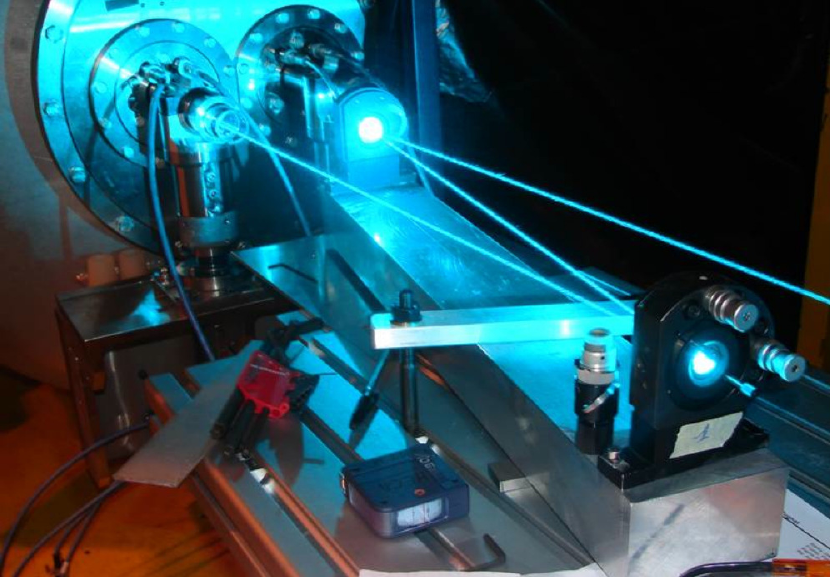}
\includegraphics[keepaspectratio=true, width=8.5cm]{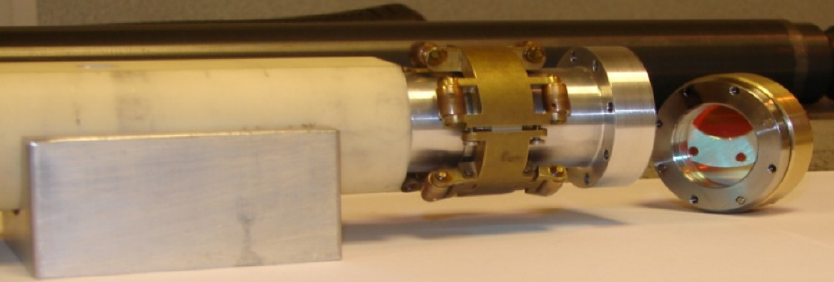}
\caption{(Upper) Intracavity laser beam of the Ar$^+$ laser operating in a Z-fold extended cavity mode of 10 m long. The output coupler of the Ar$^+$ laser, which can be seen close to one of the intracavity guiding mirrors, has been replaced by another one with a curvature radius of 10 m. (Lower) New used output coupler of 2 inch of diameter and 99.55\% of reflectivity. It has been mounted on the mole supporting structure, which has been inserted inside the magnet aperture to operate the Ar$^+$ laser in extended cavity mode.}
\label{fig:osqarlaser}
\end{figure}

The implementation of the resonantly-enhanced conversion setup is challenging with one of the main technical difficulties resulting from the design and realization of active locking systems for two high-finesse Fabry-Perot cavities of long length \cite{Bahre2013,Mueller2009}. An intermediate step considered as being necessary is currently under study. It is based on the development of a 20~m long extended laser cavity to take advantage of the large laser intracavity optical power for the ALP source. In that way, the locking of the high-finesse Fabry-Perot cavity of the photon regeneration zone to the laser in extended cavity mode will become a more affordable operation as a first step. First experimental trials with Ar$^+$ laser operating in extended cavity mode have been successful for linear as well as for Z-fold configurations of the cavity, the latter allowing better adjustments of the laser beam directions (Figure \ref{fig:osqarlaser} (Upper)). For both these configurations, the output coupler of the Ar$^+$ laser has been replaced by a mirror with a reflectivity of 99.55 \% and a radius of curvature of $10~m$. The latter has been mounted on a mole type support that can be inserted and translated inside the magnet aperture (Figure \ref{fig:osqarlaser} (Lower)).
A stable beam with an intracavity optical power of few hundred watts has been obtained. A precise measurement method of this intracavity optical power based on Rayleigh scattering \cite{Hillman1983} is under development. Among further improvements that need to be achieved it can also be emphasized the use of a mirror with higher reflectivity and a curvature radius of 20~m allowing the laser cavity to be extended over the whole length of one LHC dipole connected to its cryogenic feed box.

The resonantly-enhanced photon regeneration effect can significantly improve LSW experiments \cite{Hoogeveen1991,Sikivie2007} allowing in principle to compete and even exceed the limits provided by solar telescopes, such as CAST \cite{Zioutas2009a,Zioutas2009b,Zioutas2009c}, or haloscope \cite{Asztalos2010}. Experiments based on potential extra-terrestrial WISP sources provides today the best constraints on the coupling of WISPs to photons; however they are based on some assumptions. In solar experiments the role of the solar magnetic field that might allow the conversion of WISPs into photons at the surface of the sun is neglected \cite{Zioutas2006}, whereas for haloscope it is usually assumed that axions saturate the Milky Way's halo \cite{Asztalos2010}. Purely laboratory experiments for WISP search offer a complementary approach but their sensitivity needs to be significantly further improved. In this line, an ambitious upgrade of the ALPS experiment has been approved at DESY with as ultimate objective to surpass the CAST bounds \cite{Zioutas2009a,Zioutas2009b,Zioutas2009c} in the lower region mass {\it i.e.} a sensitivity reaching the limit g$_{A\gamma\gamma} \sim 2\times10^{-11}$~GeV$^{-1}$  \cite{Bahre2013}.

\section{Conclusion}

To conclude the obtained limits for the di-photon coupling constants of scalar and pseudo-scalar particles have confirmed the results obtained by ALPS collaboration. The mechanical stability achieved for the present 53 meter long OSQAR experiment using two spare superconducting dipole magnets of the LHC gives full confidence on the feasibility of the different anticipated upgrades.

\begin{acknowledgements}   

We thank the referees for their constructive comments. This work was partly supported by the Grant Agency of the Czech Republic 203/11/1546.

\end{acknowledgements}


\begin{thebibliography}{9}

\bibitem[] {JaRi2010} J.~Jaeckel and A.~Ringwald, Ann. Rev. Nucl. Part. Sci. {\bf 60}, 405 (2010) [arXiv:1002.0329 [hep-ph]]. 
\bibitem[] {AWGr2011} R.~Essig {\it et al.}, Report of the Community Summer Study 2013 (Snowmass) Intensity Frontier "New Light, Weakly Coupled Particles" [arXiv:1311.0029 [hep-ph]].
\bibitem[] {Weinberg1978} S.~Weinberg, Phys. Rev. Lett. {\bf 40}, 223 (1978); 
\bibitem[] {Wilczek1978} F.~Wilczek, Phys. Rev. Lett. {\bf 40} 279 (1978).
\bibitem[] {Peccei1977} R.~D.~Peccei and H.~R.~Quinn, Phys. Rev. D {\bf 16}, 1791 (1977);
\bibitem[] {Svrek2006} P. Svrek, and E. Witten, J. High Energy Phys. {\bf 06}, 051 (2006) [arXiv:hep-th/0605206].
\bibitem[] {Cicoli2012}M.~Cicoli, M.~Goodsell and A.~Ringwald, J. High Energy Phys. {\bf 2012}, (10) 146 (2012)[arXiv:1206.0819 [hep-th]];
\bibitem[] {Bradlay2003} R.~Bradlay, {\it et al.}, Rev. Mod. Phys. 75 (2003) 777-817;
\bibitem[] {Arias2012a} P.~Arias {\it et al.},  J. Cosmol. Astropart. Phys.  {\bf 06}, 013 (2012) [arXiv:1201.5902 [hep-ph]].
\bibitem[] {Arias2012b} A.~Ringwald [arXiv:1310.1256 [hep-ph]].
\bibitem[] {Meyer2013}  M.~Meyer, D.~Horns and M.~Raue, Phys. Rev. D {\bf 87}, 035027 (2013) [arXiv:1302.1208 [hep-ph]];
\bibitem[] {Melendez2012a}  B.~Melendez, M.~M.~Bertolami and L.~Althaus, Proc. of 18th Eur. White Dwarf Workshop [arXiv:1210.0263 [hep-ph]]
\bibitem[] {Melendez2012b}  J.~Isern {\it et al.}, Proc. of 8th Patras Workshop on Axions, WIMPs and WISPs [arXiv:1304.7652 [hep-ph]]
\bibitem[] {Cicoli2014}  M.~Cicoli {\it et al.} [arXiv:1403.2370 [hep-ph]]
\bibitem[] {Pugnat2010} P.~Pugnat,  {\it et al.} (OSQAR collaboration), CERN-SPSC-2010-004 / SPSC-M-770, 18 January 2010, unpublished;  
http://cdsweb.cern.ch/record/1233928/files/SPSC-M-770.pdf
\bibitem[] {Sulc2013} M. Sulc, et al., Nucl. Instr. and Meth. A 718 (2013) 530 
\bibitem[] {Cameron1993} R.~Cameron, et al., Phys. Rev. D 47 3707 (1993) 
\bibitem[] {Zavattini2006} E.~Zavattini, et al. (PVLAS collaboration), Phys. Rev. Lett. 96 110406 (2006).
\bibitem[] {Robilliard2007}  C.~Robilliard, et al., (BMV collaboration), Phys. Rev. Lett. 99 190403 (2007).
\bibitem[] {Chou2008} A. S. Chou, et al., (GammeV collaboration), Phys. Rev. Lett. 100 080402 (2008).
\bibitem[] {Afanasev2008} A. Afanasev, et al., (LIPPS collaboration), Phys. Rev. Lett. 101 120401 (2008); 
\bibitem[] {Zavattini2008} E. Zavattini, et al. (PVLAS collaboration), Phys. Rev. D 77 032006 (2008).
\bibitem[] {Pugnat2008} P. Pugnat, et al., (OSQAR collaboration), Phys. Rev. D, 78 092003 (2008).
\bibitem[] {Ehret2010} K. Ehret, et al., (ALPS collaboration), Phys. Lett. B 689 149 (2010).
\bibitem[] {Sikivie1983} P. Sikivie, Phys. Rev. Lett. 51 1415 (1983).
\bibitem[] {vanKibber1987} K. Van Bibber, N.R. Dagdeviren, S.E. Koonin, A.K. Kerman and H.N. Nelson, Phys. Rev. Lett. 59 759 (1987). 
\bibitem[] {Arias2010} P.~Arias {\it et al.}, Phys. Rev. D {\bf 82}, 115018 (2010).
\bibitem[] {Hoogeveen1991}F.~Hoogeveen and.~T Ziegenhagen, Nucl. Phys. B {\bf 358}, 3 (1991)
\bibitem[] {Sikivie2007} P.~Sikivie, D.~B.~Tanner, and Karl~van~Bibber, Phys. Rev. Lett. 98, 172002 (2007).
\bibitem[] {Bahre2013} R. B\"ahre, et al., (ALPS collaboration), Any Light Particle Search II - Technical Design Report, J. of Instr. (JINST) {\bf 8}, T09001 (2013) [arXiv:1302.5647 [physics.ins-det]].
\bibitem[] {Mueller2009} G. Mueller, P. Sikivie, D.B. Tanner and K. van Bibber, Phys. Rev. D 80, 072004 (2009).
\bibitem[] {Hillman1983} L.W. Hillman, {\it et al.}, Appl. Opt. 22, 3474 (1983).
\bibitem[] {Zioutas2009a} K.~Zioutas {\it et al.}, (CAST Collaboration) Phys. Rev. Lett. 94, 121301 (2005).
\bibitem[] {Zioutas2009b} E. Arik et al., (CAST Collaboration) J. Cosmol. Astropart. Phys. 02, 008 (2009).
\bibitem[] {Zioutas2009c} S. Aune et al., (CAST Collaboration) Phys. Rev. Lett. 107 (2011) 261302, [arXiv:1106.3919[hep-ex]].
\bibitem[] {Asztalos2010} S. J. Asztalos, et al., (ADMX collaboration) Phys. Rev. Lett. 104, 041301 (2010).
\bibitem[] {Zioutas2006} K. Zioutas, et al., J. of Phys.: Conf. ser. 39 103 (2006).

\end{thebibliography}
\end{document}